\documentstyle [12pt,preprint,eqsecnum,graphics,aps]{revtex}
\begin{document}

\draft

\title{Storage Qubits and Their Potential Implementation Through a Semiconductor Double Quantum Dot}

\author {Ehoud Pazy $^1$, Irene D'Amico $^{1,2}$, Paolo Zanardi $^{1,2}$ and Fausto Rossi $^{1,2,3}$ }
\address {$^1$ Institute for Scientific Interchange (ISI), Villa Gualino, Viale Settimio Severo 65, 
I-10133 Torino, Italy}
\address {$^2$ Istituto Nazionale per la Fisica della Materia (INFM)}
\address {$^3$ Dipartimento di Fisica, Politecnico di Torino, Corso Duca degli Abruzzi 24 -10129 Torino, Italy}

\maketitle \begin{abstract}

In the context of a semi-conductor based implementation of a quantum computer
the idea of a quantum storage bit is presented and a possible implementation 
using a double quantum dot structure is considered. A measurement scheme using
a stimulated Raman adiabatic passage is discussed. 

\end{abstract}

\section {Introduction }
\label {sec:Introduction}

Quantum systems serving as computational devices have been shown to
potentially be able to perform information processing tasks intractable for devices
relying on classical physics\cite{DiVincenzo00}. This additional computational
power provided by quantum information processing devices, i.e. quantum computers, has motivated
a large number of different proposals for possible implementations (see \cite{Divincenzo99}
and references therein). A quantum computer should be able 
to perform a large number of gating operations within the typical decoherence time.
One of the main problems with solid state quantum computer
proposals based on the charge degrees of freedom, is to find a way to overcome the ``fast'' decoherence times. 
This is a general problem: coherent quantum manipulations (gating) usually implies a 
need for strong external coupling to the qubit degrees of freedom, on the other hand strong 
coupling usually causes fast decoherence. A major step in overcoming this problem 
has been recently proposed by Biolatti et al. \cite{Biolatti00}. In their proposal they suggest
ultrafast gate operations (UGO) using laser pulses to drive energy-selected interband optical
transitions. In the UGO proposal the qubit is implemented using excitonic degrees of freedom. 
UGO is much faster than gating by time dependent electrical fields. 

Our paper is concerned with the question of finding a possible scheme 
to measure the state of a qubit in a solid state quantum computation implementation: in fact
a further requirement for a quantum computer realization is the possibility
of performing projective measurements on qubits \cite{Divincenzo99}. For the purpose of error correction
one would like to be able to perform intermediate projective measurements on single qubits, during the 
operation time of the quantum computer. For this to be possible, it is necessary to
extract the information from a qubit on a time scale shorter than the decoherence time, $T_{2}$. 
Once again one faces the
problem of short decoherence time. The problem, in this case, is even more acute since there is also 
a typical time, $T_{1}$, for the decay of the information in the qubit, due to the finite excitonic lifetime. 
Short $T_{1}$ limits the available 
time for the measurement process, even when no gating operations are being performed. There have been many 
recent 
proposals for measuring the quantum state of a solid state implemented 
qubit, using single-electron transistor (SET) \cite{Shnirman98,Korotkov99,Averin00}, tunnel junctions 
\cite{Korotkov00} and ballistic point-contact detector \cite{Gurvitz97,DAverin00}. 
These proposals as well as other models,
\cite{Aleiner97,Levinson97,Stodolsky99,Buks98} which could be adapted to measure qubits
implemented using quantum dots, involve continuous 
measurements schemes \cite{Mensky98}, i.e. schemes in which 
the current through the point-contact or SET is being continuously measured.
Regarding the UGO proposal continuous measurements schemes suffer 
major drawbacks. Since measurement induces decoherence in the measured system, in the proposed measurement schemes
there should be no net current flowing through the point-contact or SET until one decides to measure. Thus the
measurement process involves switching on of electric fields, involving again time scales which are long compared 
to the decoherence time $T_{2}$, and the excitonic life time $T_{1}$. Furthermore in these proposed measurement schemes
even when there is no net current flow through the measurement apparatus, i.e. no
electric field, still there are current fluctuations. These current fluctuations
induce a random electric potential in the qubit, i.e. decoherence.
In this paper we will show a possible way to overcome these problems via the use of a ``storage qubit''. 
First we  will introduce the idea of the storage qubit, then the measurement of the qubit by the 
storage qubit will be described and finally we will present a possible implementation of the storage qubit
using a double quantum dot (QD) structure.

\section {Storage qubit}
\label {sec:Sqb}

The idea of a storage qubit (S-qubit) is to transfer the information from the qubit to another qubit (the 
S-qubit) where the information can reside for a long time, i.e. the S-qubit posses a large $T_{1}$ 
compared to the original qubit. Moreover through the use of a S-qubit one can increase the spatial distance 
between the qubit and the measurement device, decreasing the decoherence rate when no measurement is taking place.
Due to its relatively large $T_{1}$, the information inside the S-qubit can
be extracted by the proposed continuous measurement schemes, without affecting the qubit. 
The S-qubit will measure the qubit in a time that is
``short'' compared to the decoherence time, and store the information.
The generic way to describe this measurement is through
the ``controlled not'' or c-not gate, which is also referred to as the measurement gate \cite{Paz00}. 
The measurement of the qubit by the S-qubit is thus described in the following way:

\begin{equation}
\label{eq:a1}
( \alpha | 0_{QB} \rangle + \beta | 1_{QB} \rangle) | 0_{SQB} \rangle \ \longrightarrow \ \ \alpha | 0_{QB} \rangle 
| 0_{SQB} \rangle + \  \beta | 1_{QB} \rangle | 1_{SQB} \rangle ,
\end{equation}
where $| i_{QB} \rangle$ and $| j_{SQB} \rangle$ ($i,j \epsilon \{0,1\}$),
are the qubit and S-qubit states respectively \cite{remark0}.
This sort of measurement is just the standard Von Neumann measurement model in which the time evolution 
operator is
the generator of translations in the pointer basis \cite{Paz00}, and the shift in the pointer basis 
is made accordingly to the initial state of the qubit: if the qubit is initially in state 
$| 0_{QB} \rangle$ the pointer state 
is shifted by $0$, if the qubit is initially in state $| 1_{QB} \rangle$ the pointer is shifted by $1$. 

\section {Implementation of a storage qubit}
\label {sec:Implementation}

We now consider a possible implementation of a S-qubit with the use of the double dot (DD) system described 
in a recent paper by Hohenester et al. \cite{Hohenester00}. This proposed S-qubit could be used 
for measuring the quantum state
of the qubit implementation proposed in the UGO scheme. We thus start by describing the computational 
subspace as defined 
accordingly to UGO proposal \cite{Biolatti00}. The qubit is implemented through excitonic degrees of 
freedom in a QD.
The two possible states of the qubit, $| 0_{QB} \rangle $ and $| 1_{QB} \rangle$ consist of
the absence and presence of a ground-state exciton in the QD respectively.

\begin{figure}
\scalebox{0.5}{\includegraphics[-40mm,-40mm][60mm,120mm]{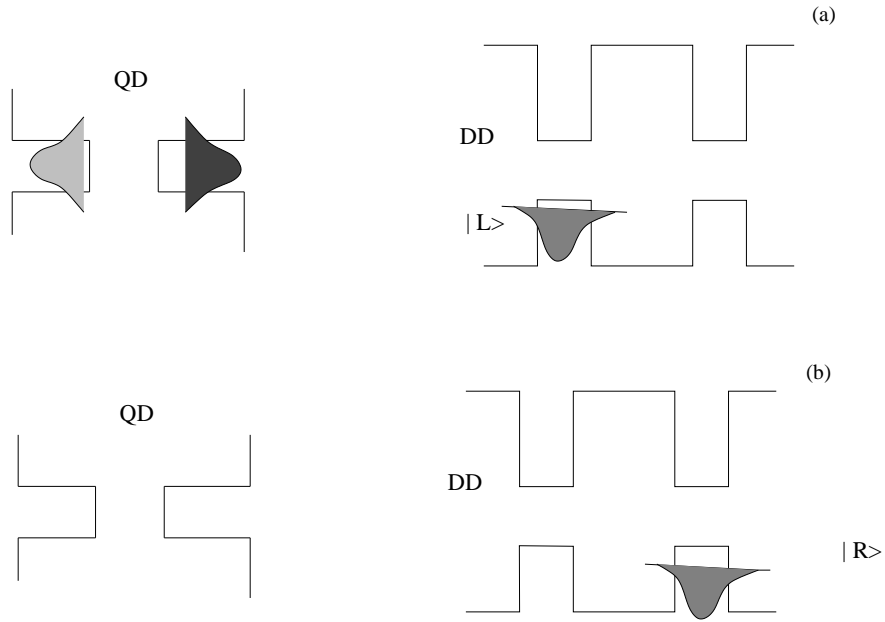}}
\caption{Schematic illustration of the implemented qubit (QD) - storage qubit (DD) structure.
The double dot states are labeled $|L \rangle$ and $|R \rangle$ and correspond to a hole
in the left dot or in the right respectively. The degeneracy between these two states is lifted
by an external electric field. (a) State of DD corresponding to an exciton in the QD the STIRAP
is detuned and the hole remains in state $|L \rangle$.
(b) State of DD corresponding to no exciton in the QD, STIRAP is not detuned and hole is transferred to
state $|R \rangle$.}
\label{fig1}
\end{figure}

The S-qubit designed to measure the excitonic state of the QD, consists of two coupled semiconductor QDs. 
Through the application of an external gate voltage a surplus hole occupies the DD 
system. The S-qubit states are thus defined as  
excess hole in right QD, $| R\rangle$ and excess hole in the left QD, $| L \rangle$ \cite{remark1}.
The original symmetry between the two states is lifted through the application of an electric field $F=15 kV/cm$
in the growth direction \cite{Hohenester00}. Due to this field the energy levels are lowered in the left dot 
with respect to the right.
For the measurement process of the qubit by the S-qubit we propose the use of coherent population transfer in
coupled semiconductor QDs, as recently proposed in Ref. \cite{Hohenester00}. The coherent 
population transfer (in this case the transfer of excess hole from the left to the right QD) is achieved 
through a Stimulated Raman Adiabatic Passage (STIRAP) \cite{Bergmann98}.  The idea is to use the Coulomb 
interaction, between the exciton in the QD and the surplus hole in the DD to detune the coherent 
population transfer in the DD (see Fig.\ref{fig1}).

For the DD to be an implementation of a S-qubit one should check the following properties: first 
the measured information about the state of the QD stored 
inside the DD should be long-lived, that is states  $| R \rangle$ and $| L \rangle$ 
should be long-lasting, i.e. the tunneling between them should occur on a much larger time scale than 
the decoherence
process $T_{2}$, and the exciton recombination time  $T_{1}$, in the QD. Second the measurement of the qubit
by the S-qubit should be fast and reliable. For the measurement to be fast, the typical time for 
extracting information
on the excitonic state of the QD should be much shorter than 
$T_{1}$ and $T_{2}$. For the measurement to be reliable, the energy shift of the DD states due to the 
existence of an exciton on the QD should be larger than the energy 
uncertainty of the laser pulses and larger than the typical width of the energy levels due to 
interaction with the environment.

It should be mentioned that in the coming sections the estimates presented are based on the same parameters
used in Ref.\cite{Hohenester00}, except for the distance between the two QDs which has been extended to 
$100 \AA~$. This change of the distance between the dots will be discussed extensively 
when describing the measurement scheme.

\subsection {Estimation of storage life time}
\label {subsec:Lifetime}

We begin by showing that the states $| R \rangle$ and $| L \rangle$ are long lived. 
A rough estimate \cite{remark3} of the tunneling rate, $\nu$, between these two states can be given by their overlap 
times an attempt frequency ($\nu_{0}$),
$\nu = \nu_{0} \exp \left( -2 r / \xi \right)$, where $r$ is the distance between the two
QDs and $\xi$ is the localization length. 

$\nu_{0}$ is of the order of several pico-seconds and it
can be approximated by $\nu_{0} \approx {\hbar \over 2 m_{h} l^2}$, where 
$l = 50\AA~$  is the well width. The localization length can be estimated by 
$\xi \approx {\hbar \over \sqrt{2 m_{h} (V-E)}}$, where $V-E = 200 \ meV$ 
is the effective potential barrier between the two hole states and $m_{h}= 0.34 m_{0}$ ($m_{0}$ is the 
free electron mass) is the hole mass. 
Taking the distance between the two QDs to be $r = 100 \AA~$, one gets a tunneling time of
the order of one nano-second. The naive approximation for the tunneling time between the two
DD states, $| R \rangle$ and $| L \rangle$, is orders of magnitude larger than the typical time 
$T_{2}$, for the QD exciton state. 

In a more refined estimate of the tunneling time between the
two hole states, one has to consider the effects of the coupling to the phonon environment.
The major effect of the coupling to phonons is an activational process.
As mentioned there is an applied external electric field in the growth direction. Due to this field the energy 
levels are lowered 
by $\Delta E \approx 20 meV$ in the left dot with respect to the right. Therefore there is a 
further exponential reduction term $\exp({-\Delta E \over  k_{B} T})$, where 
$T$ is the temperature and $k_{B}$ is the Boltzmann constant, 
when tunneling from $| L \rangle$ to $| R \rangle$. This factor is
due to the fact that the tunneling is inelastic and one has
to consider the probability of absorbing a phonon of energy $\Delta E$.  
Through this activational factor, for temperatures of the order of a few Kelvin, the time information is stored in 
state $| L \rangle$ is extremely long, and the transition between  $| L \rangle$ to  $| R \rangle$ is highly 
improbable \cite{Caldeira84}. 

\subsection{Measurement using a STIRAP process} 
\label{subsec:measurement}
Before describing the proposed measurement process we give a short description of the STIRAP process in the DD
structure. The STIRAP process consists of three states, 
two of which are the long-lived lower energy states $| R\rangle$ and $| L\rangle$, between 
which there are no dipole allowed 
transitions. Both these levels ( $| R\rangle$ and $| L\rangle$) are instead dipole coupled to a third, 
higher energy state, in 
this case a charged exciton state labeled  
$| X^{+}\rangle$. Through the use of two delayed laser pulses coherent population transfer can be achieved between 
$| L \rangle$ and $| R \rangle$ without ever occupying state $| X^{+} \rangle$. The first pulse (``Stokes''), is
tuned to the $R$-$X^{+}$ resonance and the second pulse (``pump'') is tuned to the $L$-$X^{+}$ resonance. 

\begin{figure}
\scalebox{0.5}{\includegraphics[-80mm,0mm][80mm,80mm]{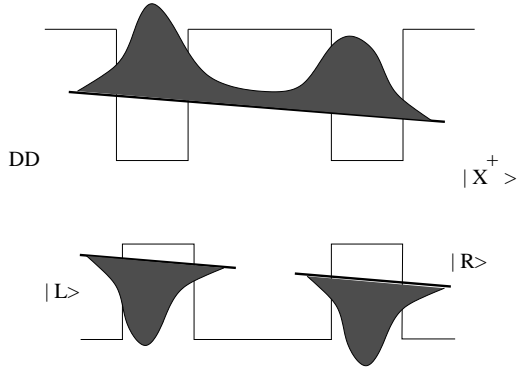}}
\caption{Schematic illustration of the charged exciton state in the double dot structure, $| X^{+} \rangle$.
The two holes are in their ground states, while the electron is an excited state such that its wave function
is split between the QDs.}
\label{fig2}
\end{figure}

For the STIRAP process to be effective the coupling of the excited state $| X^{+} \rangle$ 
to the two long lived states should be of the same order. Moreover the two long lived states 
should be non-degenerate. In Ref.\cite{Hohenester00} this is achieved first by an electric 
field (in the growth direction) which lifts the original degeneracy of states  $| R\rangle$ and 
$| L\rangle$ and second, whereas the hole states are localized, the electron 
wave function in the excited state $| X^{+} \rangle$ is split between the two QDs. 
This splitting of the electron wave function allows the coupling between the $| X^{+} \rangle$ state to the two 
states $| R\rangle$ and $| L\rangle$ to be of the same order. In our proposed scheme for an implementation 
of a S-qubit, we have increased the parameter for the spatial separation between the two wells to $r = 100 \AA~$.
This localizes the ground state and the first excited states of the electron
in one of the QDs. To have an electron wave function \cite{remark4} 
 which is spread in a similar way over the two QDs, which is needed for an effective STIRAP process, one can think
of two options. The first is using a charged exciton in which the holes are in the ground states in both QD and the
electron is in a high energy level in the QD, i.e. comparable to the confining potential. Thus in this case the
proposed $| X^{+} \rangle$ excited state of our implementation scheme is composed of two localized
hole functions in the two QDs and an electron wave function which is split between the wells
(see Fig.\ref{fig2}). 

A second possibility is to have the holes again in their ground state and the electron excited to a continuum level
above the QD confining potential. In this case the charged exciton state is a hybrid state
of a confined exciton state for the hole and a bulk exciton state for the electron. The typical length
scales for the hole wave function are given in this case by the confining potential width, $l = 50\AA~$ and for the 
electron by the Bohr radius, $a_{B}=95\AA~$. Both the above possible excited states for $| X^{+} \rangle$ are 
very susceptible to decoherence. Especially
the hybrid state where the electron is in a continuum level bound to the hole by Coulomb interaction is prone
to decoherence: in fact outside the QD the electron is not shielded by the QD confining potential from interacting with 
phonons or other decoherening mechanisms.

The measurement of the QD state (exciton or absence of exciton in the QD) is based on exploiting the
Coulomb interaction between the exciton and the charged states in the DD, i.e.
hole in left (right) dot and charged exciton. The idea is to use the shift of the energy levels in the DD
due to Coulomb interaction with the QD, to detune the coherent 
population transfer, in a way similar to what is done in the c-not gating operation in Ref.
\cite{Biolatti00}. The presence of an exciton in the QD prevents the coherent transfer of
the excess hole from the left QD to the right QD of our DD by detuning the STIRAP process (see Fig.\ref{fig1}).

Concerning decoherence, one
requires from the measurement device, i.e. the DD, to not decohere the QD
when no measurement is taking place. This requirement is fulfilled since
the presence of the hole in the DD apparatus does not disturb the QD states 
rather it causes a constant (time independent on the scale of the computation time) shift of the 
energy levels. Thus the measuring device will not affect the quantum computer 
when the measurement is not taking place.

Regarding the typical time on which the measurement takes place,
the measurement of the state of the QD by the DD occurs on a time scale which is given by the duration
of the laser pulses ``Stokes'' (``pump'') which induce the coherent population transfer. The duration of the 
laser pulses is of the order of $10 \ ps$ \cite{Hohenester00}. Thus the typical time for extracting 
information on the state of the QD is fast
compared to the excitons dephasing and recombination times (the dephasing time being of the order of $100 \ ps$ 
\cite{Hohenester00}). 

\subsection{Shift of the energy levels of the double quantum dot} 
\label{subsec:shift}

The measurement process of the QD by the DD is done via the detuning of the  STIRAP process. The STIRAP 
process is rather robust: since in the adiabatic limit its efficiency is unaffected by the perturbations 
of the virtual intermediate state and also it lacks sensitivity to small detuning of this state, 
still it is susceptible 
to detuning. In order for the STIRAP to take place one needs the adiabatic condition to be fulfilled, i.e. 
$\Omega \tau \gg 1$, where $\tau$ is the duration of the pulses overlap and $\Omega$ is the typical Rabi frequency
associated with the STIRAP process. A much stronger constraint is that the initial and final levels have to be in 
resonance in order to fulfill the energy conservation requirement during the transfer \cite{Elk95}. For the 
measurement process we need the STIRAP to take place {\it only} when there is {\it no} exciton in the QD.

We shall now show how the existence of an exciton destroys the probability for a STIRAP process to take place by,
first, detuning the intermediate level such that the adiabatic condition is not fulfilled, and, more important, by 
moving the final and initial levels out of resonance. When an (energy) detuning $\Delta_{p}$ of the pump laser 
from resonance with the $L$-$X^{+}$ transition and a detuning of the Stokes laser from the $R$-$X^{+}$ transition
$\Delta_{s}$ are introduced, the Hamiltonian for the three-level system within the rotating
wave approximation introducing has the form \cite{Bergmann98}
\begin{equation}
\label{eq:Hamiltonian} 
H = {\hbar \over 2}\left [ ( \ \Omega_{p} | X^{+} \rangle \langle L| + \Omega_{s} | X^{+} \rangle \langle R| + h.c\ )
+2 \Delta_{p} | X^{+} \rangle \langle  X^{+}| + 2(\Delta_{p} - \Delta_{s}) |R \rangle \langle R| \right ]
	 \ ,
\end{equation}
where $\Omega_{p}$ and $\Omega_{s}$ are the coupling Rabi frequencies, corresponding to the pump and Stokes respectively.

We first consider the case when the two photon resonance condition applies, i.e. $ \Delta_{p}=\Delta_{s}$. 
The instantaneous eigenstates and eigenfunction of the Hamiltonian Eq. \ref{eq:Hamiltonian} are given by
\begin{equation}
\label{eq:eigen}
\begin{array}{lcl}
|a_{0} \rangle   = & \cos \theta |L\rangle - \sin \theta |R\rangle, & \  \ \ \omega_{0} =  0 \nonumber \\
|a_{\pm} \rangle  \propto & \sin \theta |L\rangle \pm \cot^{\pm 1} \phi | X^{+} \rangle + \cos \theta |R\rangle, & \ \ \ 
\omega_{\pm}  =  \Delta_{p} \pm \sqrt{\Delta_{p}^2 + \Omega_{p}^2 + \Omega_{s}^2} ,
\end{array}
\end{equation}
where $\theta$ is the mixing angle defined by, $\tan \theta ={\Omega_{p} \over \Omega_{s}}$ and $\phi$ is
given by the detuning and Rabi frequencies and is of no importance in the ensuing discussion. $|a_{0} \rangle$
is referred to as the ``dark state'' since it includes no contributions from the ``leaky state'' $| X^{+} \rangle$.

The condition for an adiabatic transfer is given by $|\omega^{\pm}-\omega_{0}| \tau \gg 1$. 
For the parameters used in the 
Hohenester et al. \cite{Hohenester00} paper ($\Omega_{s,p}=1.0 \ meV, \ \tau=10 \ ps$) when the the laser detuning 
$\Delta_{p}$ becomes of the order of the effective Rabi frequency, $\Omega_{eff}=\sqrt{\Omega_{p}^2+\Omega_{s}^2}$, 
the adiabatic condition is no longer fulfilled. In this case the STIRAP process is detuned when the levels in the
DD are shifted such that the energy difference for the transition $L$-$X^{+}$ is shifted by more than $1.0 \ meV$.
When the adiabatic condition is no longer fulfilled one has a non-vanishing probability for occupying the leaky state.
Once the leaky state is occupied there is high probability of a transition to a different state, i.e. not one of the 
three states used for the STIRAP process. In this way the hole transfer from the left QD to the right does not take 
place.

\begin{figure}
\scalebox{0.5}{\includegraphics[-20mm,0mm][80mm,120mm]{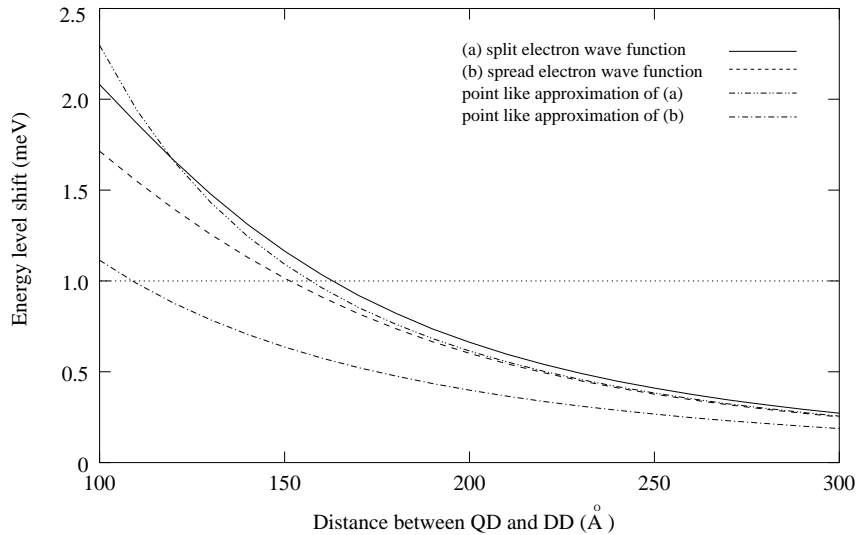}}
\caption{Shift of energy levels of the DD as a function of the distance from the QD. The distance is measured
from the center of the electron wave function in the QD to the center of the left (closest) QD of the DD 
configuration. For the case of an excited electron state spread over the dots we took the typical 
length scale for the wave function, in the growth direction, to be $100\AA$. 
Results are presented for Gaussian and ``point like'' wave functions for two cases: 
electron wave function split between the QDs
and electron wave function spread over the two QDs.}
\label{fig3}
\end{figure}

As described in the Appendix, we have estimated that, when the electron wave function of the excited 
exciton state is split between the two QD, then up to a distance of
$170 \AA~ $ between the QD and DD, the energy level shift in the DD due to the presence of an exciton in the 
QD is bigger than $1.0 meV$ (see Fig.\ref{fig3}). It is worthwhile to note that using an excited state which is a 
hybrid between a bulk exciton for the electron and a confined exciton for the hole we obtain an energy level shift 
in the DD bigger than $1.0 meV$ for distances up to $150 \AA~ $. It is therefore not crucial to get the excited 
electron localized inside the QDs, since from our estimates any excited electron state can provide the needed 
energy shift for 
distances up to $150 \AA~ $. It is also interesting to note that the excited electron state which is split 
between the two dots, gives an energy shift which is quite similar to what one would obtain by using two 
point like charges (each of charge $ {e \over 2}$) sitting in the center of the two dots (see Fig.\ref{fig3}). 
This means that the point like approximation is quite good for the localized excited electron state.

\begin{figure}
\scalebox{0.55}{\includegraphics[-20mm,0mm][120mm,100mm]{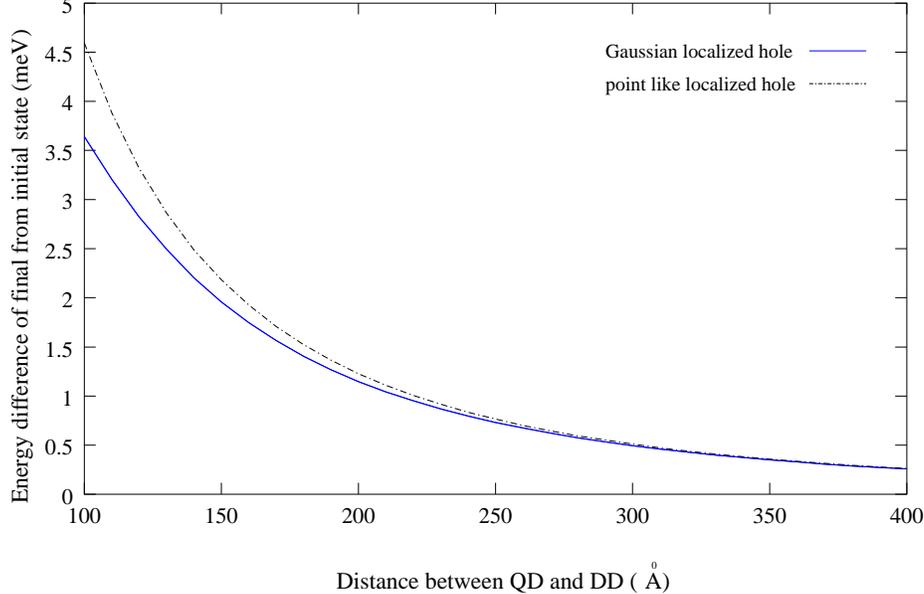}}
\caption{Difference of the shift of initial state, $|L \rangle$ and the final state, 
$|R \rangle$, of the STIRAP process in the DD when there is an exciton in the QD as a 
function of the distance from the QD (the distance is measured in the same way as described 
for Fig \ref{fig3}). Results are presented for Gaussian and ``point like'' wave functions.}
\label{fig4}
\end{figure}

For the case when $ \Delta_{p}\ne \Delta_{s}$, the STIRAP process is destroyed much sooner. Taking even a small
non zero $ \epsilon \equiv \Delta_{p}- \Delta_{s} \ll 1$ one does not get a dark state any more. The zero
eigenvalue moves to a value of the order of $\tilde{\omega}_{0} \simeq {2 \Omega_{p}^2\ \epsilon \over \Omega_{eff}^2}$ 
which changes the dark state, $|a_{0} \rangle$, such that it includes contributions from the leaky state,
$|X^{+}\rangle$, which are of the same order as $\tilde{\omega}_{0}$. Since in this case the energy conservation 
requirement is not fulfilled, i.e. the final and initial levels are not in resonance, in order to see if the 
STIRAP process takes place
one needs to compare the energy uncertainty of the pulse with the difference in the energy shift
of the initial and final states. Therefore the condition for the STIRAP to take place is given by
$\epsilon \tau \leq 1$ \cite{remark5}.  
Since $\tau= 10 ps$ it would be enough to have the energy shift larger than $0.5 meV$. In 
Fig.\ref{fig4} we show the difference in energy between the initial state, $|L \rangle$ and the final state, 
$|R \rangle$, of the DD when there is an exciton in the QD. This energy difference is much more susceptible to
the existence of an exciton in the QD and is shown to be greater than $0.5 meV$ up to distances of $300 \AA~$ 
between the QD and DD centers (for details of the calculation see the Appendix).  

\noindent The reason for which it is much easier to detune the transition with respect to the initial and 
final states (hole in $|L \rangle$ , hole in $| R \rangle$) rather than the initial (or final) 
state with regards to the intermediate state is due to the different charge configurations. The intermediate, 
leaky, state ($|X^{+}\rangle$) couples in a weaker way to an exciton in the QD since the detuning is basically 
due to a dipole-dipole interaction, whereas for the initial and final states the detuning is due to a 
charge-dipole interaction.
\newpage
A possible alternative measurement scheme is the use of a two $\pi$ pulse process. In this process one
would first excite the system from the initial state to the intermediate, leaky, state via a $\pi$
pulse then using a further $\pi$ pulse drive the system from the intermediate state to the final state, i.e.
hole in right QD. There are two strong arguments in favor of using the STIRAP over this alternative idea.
First of all in order to use the two $\pi$ pulse method it is necessary to
detune it, i.e. detune one of the two transitions
involving the intermediate state. Since the coupling of this state to the exciton state in the QD is weaker,
this method of implementing a measurement would be less effective. Second in variance to the STIRAP process
using the two $\pi$ pulse process would involve the occupation of the intermediate, leaky, state, thus involving
unwanted transitions, i.e. losses.  

\section {Summary}
\label {sec:Summary}
To summarize, we have provided a measurement scheme which is fast compared to the short decoherence times typical of 
semi-conductor implementation schemes of a quantum computer. We have proposed the idea of a storage qubit which 
is used to measure the quantum computer's qubit state and to store the information for a longer time so 
it can be extracted using conventional methods, e.g. using a SET 
\cite{Shnirman98,Korotkov99,Averin00} or a point contact \cite{Gurvitz97,DAverin00}. 
A possible implementation of a storage
qubit using the stimulated Raman adiabatic passage process was presented. This implementation scheme was shown to 
fulfill all the necessary requirements for a storage qubit: the information is stored in the storage qubit for a 
much longer time scale than in the qubit, and it is possible to perform a fast and reliable measurement of the 
qubit by the storage qubit. 

There are some issues concerning the scalability of the specific implementation scheme suggested
for a storage qubit. One could argue that having a double quantum dot as a small measuring device for each
quantum computer qubit might constitute a too strong constraint on the scalability of the quantum computer. 
This constraint on scalability is twofold: first one might question whether it is possible to arrange 
geometrically all these quantum dots and double quantum dots such that each quantum dot has a neighboring 
double dot and still have the possibility for an operating quantum computer. 
Secondly one might remark that qubits are difficult enough to construct and having such a 
measurement scheme enlarges the difficulty by at least a factor of two. These arguments are not 
overwhelming: first of all it is possible to conceive a geometrical configuration in which the quantum computer 
is arranged on a plane while the storage double dot structures are located in planes above and below it. 
Another possibility is to consider measuring only certain qubits in the quantum computer i.e. specific quantum 
dots. The information can be transferred to these measurable quantum dots in a cellular automaton sort of scheme.

\section{Acknowledgments}
We are grateful to P. Zoller, U. Hohenester, E. Biolatti and R. Ionicioiu for fruitful discussions. 
This work was supported in part by the European Commission through the Research Project {\em SQID} within the
{\em Future and Emerging Technologies} (FET) programme.

\newpage

\appendix

\section*
{Estimate of the shift of the transition frequency in the DD apparatus}

In the appendix we first calculate the detuning of the pump laser from resonance with the $L$-$X^{+}$
transition, due to the shifting of the energy levels in the DD
and then the difference in the energy shift of the initial and final states, both due to 
the presence of an exciton in the "computing" QD (CQD). 
We begin by discussing the relative
position of the two structures, the CQD and the DD:
the presence of the hole state $|L>$ in the DD, will modify, through Coulomb interaction, 
the length of the dipole in the CQD. In this respect, growing the DD on the QD
substrate in the direction opposite to the field, i.e. with 
the DD hole aligned and closer to the CQD
electron than to the CQD hole, will induce a larger dipole in the CQD 
keeping the external field unchanged.
 This would affect in a positive way the quantum computing process since it 
could be used to enhance the biexcitonic shift between excitons in
different CQD's and a large biexcitonic shift is at 
the core of the quantum computing scheme proposed in
Ref.\cite{Biolatti00}. Therefore this is the arrangement we will consider in
our calculations. 

Since both structures (DD and CQD) are in the 
strong confinement regime, i.e. the typical length scale associated to the
harmonic potential  is much smaller than the effective Bohr radius, 
we can assume for $| X^{+} \rangle$ the factorized form
$| X^{+} \rangle = | L \rangle | X \rangle $, where 
$\langle {\bf r}| X \rangle = \psi_{e}^{dd}({\bf r}) \psi_{h}^{dd}({\bf r}) $ 
consists of a ground state hole in the right QD ($\psi_{h}^{dd}({\bf r})$) and an excited electron wave function
$\psi_{e}^{dd}({\bf r})$. Similarly the exciton wave function
in the CQD structure will be factorized as $\psi_{exc}({\bf r})=
\psi_{e}^{qd}({\bf r})\psi_{h}^{qd}({\bf r})$, where  $\psi_{h(e)}^{qd}({\bf r})$ is
the hole (electron) single particle wave function.

It should be noted that the single particle wave functions we have defined are obtained solving the 
Schr\"{o}dinger equation including the Coulomb interaction. As a first approximation the effects
of Coulomb interaction on bounded, low energy states, i.e. the hole states in the DD and on the states in CQD, 
can be neglected. This is not the case for the {\it excited} electron wave function 
$\psi_{e}^{dd}({\bf r})$, whose shape is definitely influenced by the Coulomb interaction with the two holes.

Due to the factorization described above, the transition frequency shift will be given by the Coulomb interaction
between the CQD exciton and $| X \rangle $, i.e. one does not need to consider the change in the left hole state
$| L \rangle$ due to the state $| X \rangle $.
The expression for the energy shift becomes:
\begin{eqnarray}
\label{eq:ap1}
\Delta E & = & \Delta E_{X^+} -\Delta E_{L} \nonumber \\& = & {e^2\over\epsilon} \int d^3 r_{1} \int d^3 r_{2} \int d^3 r_{3} \int d^3 r_{4} \
{\mid\psi_{e}^{qd}({\bf r_1})\mid}^2 {\mid \psi_{h}^{qd}({\bf r_2})\mid}^2 {\mid \psi_{e}^{dd}({\bf r_3})\mid}^2
{\mid
 \psi_{h}^{dd}({\bf r_4})\mid}^2 \nonumber \\
& \times & \left ( {1 \over |{\bf r_1} - {\bf r_3}|} - {1 \over |{\bf r_2} - {\bf r_3}|} -
{1 \over |{\bf r_1} - {\bf r_4}|} + {1 \over |{\bf r_2} - {\bf r_4}|} \right ),
\end{eqnarray}
where $ \Delta E_{X^+}$ and $\Delta E_{L}$ are, respectively, 
the energy shifts of states $| X^{+} \rangle$ and $| L \rangle$ due to the existence of an exciton in the CQD.

To calculate the expression (\ref{eq:ap1}), we assume a three dimensional 
Gaussian form for the ground state single particle wave functions
$\psi_{e}^{qd}$, $\psi_{h}^{qd}$ and $\psi_{h}^{dd}$. In the quantum dot plane, 
in which the confining potential is modeled as a harmonic potential of frequency $\hbar\omega_i^j$, $i=e,h$ and 
$j=qd,dd$, their width is given, as expected, by
$\lambda_i^j=\sqrt{\hbar/m_i\omega_i^j}$. The values used are $\hbar\omega_{e}^{qd(dd)} = 30 meV$ for the electron states 
(both in the CQD and the DD) and $\hbar\omega_{h}^{qd} =24 meV$ for the hole state in 
the CQD and $\hbar\omega_{h}^{dd} = 5meV$ for the hole
state in the DD  \cite{remark2}. In the growth direction $z$, in which the potential is modeled as a 
square well large
$a$, the width is taken to be $\sqrt{\langle z^2\rangle}$, where the average is done over the ground state of the
corresponding infinite square well. Replacing the wave functions in the $z$ direction by Gaussians in this way
simplifies the calculations. This is 
a good approximation, since the difference between the two functions, i.e. single particle wave function and 
Gaussian approximation, is very small.
For the excited, delocalized, state $\psi_{e}^{dd}$  we consider two different possibilities, the first
corresponding to a state only weakly bounded (by the confining potentials) in the $z$ direction, 
the second to a state bounded (only due to Coulomb interaction) in the $z$ 
direction. In the first case, $\psi_{e}^{dd}$  is modeled as the sum of two Gaussians, each on them 
centered in one of the two dots of the DD structure. 
Their width in the in-plane directions is still given by the confining harmonic potential,
i.e. ${\lambda_i}^{dd}= \sqrt{\hbar/m_i{\omega_i}^{dd}}$, while in the growth direction (instead of using
$\sqrt{\langle z^2\rangle}$) it is simply  given  by the 
box size $a$. This choice accounts both for the wider spreading of the excited state and for the fact 
that the state is still confined by the DD wells.
In the second case, the in-plane structure of $\psi_{e}^{dd}$  remains the same, while
its $z$ component is a Gaussian centered in the middle of the DD structure and of width
$100\AA$, i.e. roughly the effective Bohr radius of the material.

In this approximation, the expression (\ref{eq:ap1}) can be reduced to 
a sum of two dimensional integrals, which are numerically easy to calculate. 
The calculation of the transition frequency shift (see Fig. \ref{fig3}) shows that, 
independently of the choice for $\psi_{e}^{dd}$, $\Delta E\approx 1 meV$ when the 
distance between the two structures CQD and DD is of the order of $150\AA$.  

The calculation of the difference in energy shifts of the states $| L \rangle$ and $| R \rangle$ is much simpler,
since one has only to calculate the following expression 
\begin{eqnarray}
\label{eq:energylr}
\Delta E & = & \Delta E_{L} -\Delta E_{R} \nonumber \\
& = & {e^2\over\epsilon} \int d^3 r_{1} \int d^3 r_{2} 
\int d^3 r_{3}\
{\mid\psi_{e}^{qd}({\bf r_1})\mid}^2 {\mid \psi_{h}^{qd}({\bf r_2})\mid}^2 
\left ( {\mid \psi_{h}^{dd}({\bf r_3})\mid}^2 - {\mid \tilde{\psi}_{h}^{dd}({\bf r_3})\mid}^2 \right ) \nonumber \\
& \times & \left ( {1 \over |{\bf r_1} - {\bf r_3}|} - {1 \over |{\bf r_2} - {\bf r_3}|} \right ),
\end{eqnarray}
where $\tilde{\psi}_{h}^{dd}({\bf r})$  is the ground state hole wave function in the right QD.
Using the previous approximations, we obtain the detuning of the final and initial states presented in Fig.\ref{fig4}
as a function of the distance between the CQD and DD.


\begin{thebibliography}{aaaaaaaaaaaaaaaaaaaaa}
\bibitem{remark0}
It should be stated that this measurement does not suffer from the ambiguity of the ``system/apparatus'' which 
appears in the general c-not operation due to the possibility of the Hadamard transformation (see Ref. \cite{Paz00}),
since we require the S-qubit to be initially in the ground state.
\bibitem{remark1}
We try through the paper to keep the notations as close as possible to the referred papers: Ref. 
\cite {Biolatti00} and Ref.\cite{Hohenester00}, in order to avoid confusion.
\bibitem{remark3}
In this estimate we consider the one dimensional tunneling problem of a hole tunneling between two states defined by the 
potential in the growth direction.
\bibitem{remark4}
We are in the strong confinement regime (the confining potential width is much smaller than the Bohr radius)
thus it is meaningful to describe the state in terms of an electron and hole wave functions, rather than 
using an excitonic wave function.
\bibitem{remark5}
Making the pulse overlap short does not help since the adiabaticity condition
should also be maintained, combining the two conditions one get ${\epsilon \over \omega^{\pm}} \ll 1$.
\end{thebibliography}
\end{document}